\documentclass[aps,prl,twocolumn,superscriptaddress]{revtex4-2}

\usepackage{amsmath}
\usepackage{amssymb}
\usepackage{graphicx}
\usepackage{color}
\usepackage[colorlinks,citecolor=blue,linkcolor=blue,urlcolor=blue]{hyperref}

\newcommand{\dd}{\mathrm{d}}

\newcommand{\mbf}[1]{\mathbf{#1}}

\newcommand{\mr}[1]{\mathrm{#1}}

\begin{document}
\title{Deconfined Boundary Phase Transition of a Quantum Critical Heisenberg Model}

\author{Chengxiang Ding}
\email{dingcx@ahut.edu.cn}
\affiliation{School of Microelectronics \& Data Science, Anhui University of Technology, Maanshan 243002, China}

\author{Long Zhang}
\email{longzhang@ucas.ac.cn}
\affiliation{Kavli Institute for Theoretical Sciences and School of Quantum, University of Chinese Academy of Sciences, Beijing 100190, China}

\date{\today}

\begin{abstract}
We investigate the boundary phases of a (2+1)-dimensional quantum critical Heisenberg model with a dangling spin chain. By introducing a multispin $Q$-term along the boundary, we drive a continuous boundary transition from an antiferromagnetic (AF) order to a valence-bond solid (VBS) order. Using large-scale quantum Monte Carlo simulations, we locate the critical point at $Q_c=0.310(11)$, and obtain the critical exponents at $Q_{c}$, including $y_{s}=0.81(4)$ and the scaling dimensions of AF and VBS order parameters $\Delta_s=0.660(15)$ and $\Delta_v=0.204(14)$. The weak long-range AF order for $Q<Q_{c}$ is stabilized by quasi-long-range effective interactions mediated by the critical bulk state, while the VBS phase restores the ordinary critical behavior. Our findings highlight the synergy between topological terms and quasi-long-range interactions in low-dimensional quantum many-body systems.
\end{abstract}

\maketitle 

{\it Introduction.---}
The physical properties at the boundary of a critical system are in general different from the bulk state~\cite{Binder1983phase}. For a given critical bulk state, the boundary critical behavior can be changed by tuning the interactions on the boundary, resulting into a rich boundary phase diagram. The boundary degrees of freedom acquire quasi-long-range (qLR) effective interactions by coupling to the critical bulk state, hence it can host novel states and phase transitions that cannot be realized in isolated low-dimensional systems only with local interactions.

The boundary critical behavior of the three-dimensional (3D) O($N$) model has attracted revived interest recently~\cite{Zhang2017, Ding2018, Weber2018, Weber2019a, Zhu2021, Weber2021, Zhu2021b, Wang2022, Wang2023d, Ding2023, ParisenToldin2023, ParisenToldin2021, Hu2021, ParisenToldin2022, Zhang2022, Zhang2023a, Zou2022, Sun2022, Sun2022b, Jian2021, Metlitski2022, Padayasi2022a, Song2025}, initially motivated by questions regarding the stability of gapless boundary states of 2D topological phases against quantum phase transitions~\cite{Grover2012a, Suzuki2012b, Zhang2017}. Nonordinary boundary critical behavior was observed numerically at the (2+1)D O(3) quantum critical point (QCP) and was attributed to the coupling of a dangling spin chain with the critical bulk state~\cite{Zhang2017, Ding2018, Weber2018}.

Different theoretical scenarios addressing the boundary phases have been proposed subsequently. Metlitski~\cite{Metlitski2022} showed that a 2D nonlinear $\sigma$ model (NL$\sigma$M) without topological terms coupled with the 3D bulk state is driven into an extraordinary-log phase, which is characterized by the logarithmically decaying spin correlation function, $C_{\parallel}(r) \propto (\ln (r/r_{0}))^{-q_{\parallel}}$. This has been numerically observed on the boundary of 3D classical O($N$) ($N=2,3$) models~\cite{ParisenToldin2021, Hu2021, ParisenToldin2022, Zhang2022, Zhang2023a, Zou2022} and (2+1)D Bose-Hubbard model~\cite{Sun2022, Sun2022b}. On the other hand, Jian et al~\cite{Jian2021} adopted the SU(2)$_{1}$ Wess-Zumino-Witten model to describe the dangling spin-$1/2$ Heisenberg chain, and argued that the spin chain coupled with the bulk acquires a long-range antiferromagnetic (AF) order~\cite{Song2025} and can undergo a continuous transition to a valence-bond solid (VBS) order. This scenario incorporates the essential topological $\theta$-term in the NL$\sigma$M description of the dangling spin chain~\cite{Haldane1985, Haldane1988}; therefore, it is not ruled out by the observation of extraordinary-log behavior in classical spin models and boson models, and deserves further examination.

\begin{figure}[!tb]
\centering
\includegraphics[width=\columnwidth]{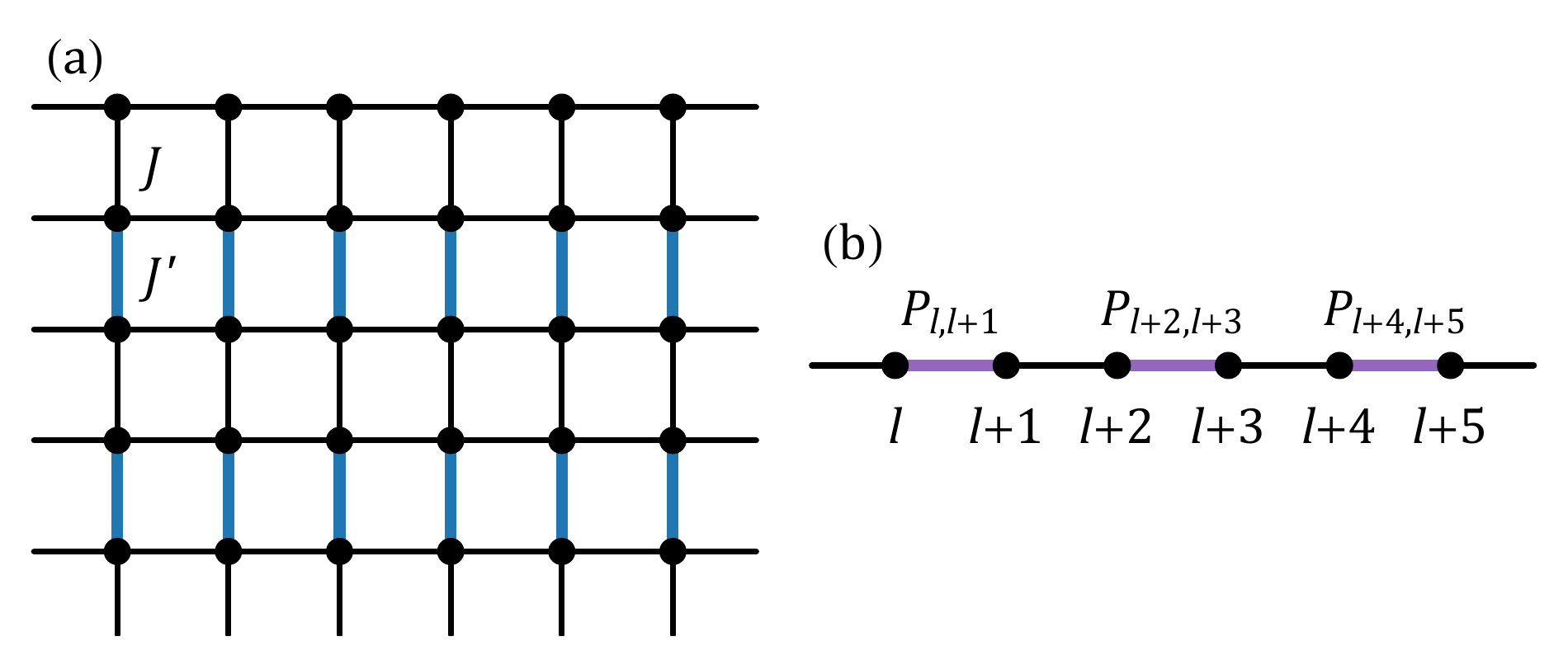}
\caption{Model. (a) Spin-1/2 Heisenberg model on the columnar dimerized square lattice. The AF interaction strengths on the weak (black) bonds and the strong (blue) bonds are $J$ and $J'$, respectively. The lattice has open boundaries with dangling spin chains in one direction, and periodic boundary condition in the other direction. (b) Illustration of the multispin $Q$-term defined in Eq.~(\ref{eq:qterm}) along the dangling spin chain.}
\label{fig:model}
\end{figure}

\begin{figure*}[tb]
\centering
\includegraphics[width=\textwidth]{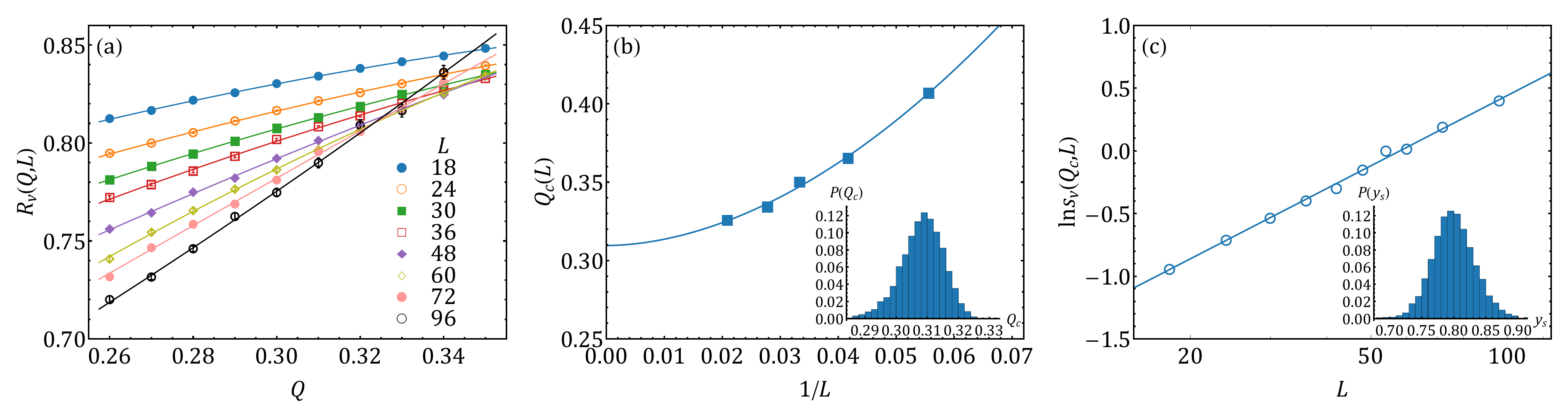}
\caption{Boundary critical point. (a) Binder ratio $R_{v}(Q,L)$ of the boundary VBS order parameter versus the tuning parameter $Q$ for different lattice sizes. The solid curves are polynomial fitting. The intersection of these curves indicates a boundary transition. (b) The crossing point $Q_{c}(L)$ versus $1/L$. The extrapolation according to Eq.~(\ref{eq:qc}) (solid curve) yields an estimate of the critical point $Q_{c}$. Inset: Histogram of estimated $Q_{c}$ values from the resampling procedure. The standard deviation quantifies the statistical fluctuations. (c) The slope $s_{v}(Q_{c},L)$ versus $L$ in the log-log scale. The solid line is the power-law fitting. Inset: Histogram of estimated $y_{s}$ values from the resampling procedure.}
\label{fig:binder}
\end{figure*}

In this work, we study the boundary phases of the (2+1)D quantum critical Heisenberg model with a dangling spin chain, which is illustrated in Fig.~\ref{fig:model}(a). A multispin $Q$-term defined in Eq.~(\ref{eq:qterm}) is introduced along the spin chain. It stabilizes a VBS order on the boundary for $Q > Q_{\mr{c}}$ with $Q_{\mr{c}} = 0.310(11)$. We identify a weak long-range AF order on the boundary for $Q < Q_{\mr{c}}$, which vanishes continuously as $Q\to Q_{\mr{c}}^{-}$. Therefore, a continuous AF-VBS transition takes place on the boundary, which is consistent with the scenario proposed by Jian et al~\cite{Jian2021}, and highlights the significant role of the topological $\theta$-term in describing the quantum spin chain coupled with the bulk state. The AF order spontaneously breaks the spin rotation symmetry, while the VBS order breaks the lattice translation symmetry; hence, the continuous AF-VBS transition is incompatible with the traditional Landau theory of phase transitions. Moreover, the low-energy excitations in the AF phase are spin waves, while excitations in the 1D VBS phase are deconfined spinons on the VBS domain walls~\cite{Tang2011a, Tang2015}. Therefore, the boundary hosts a deconfined AF-VBS transition that cannot be realized in an isolated 1D system with local interactions.

{\it Model and method.---}
We study the spin-$1/2$ AF Heisenberg model on the columnar dimerized square lattice illustrated in Fig.~\ref{fig:model}(a). The Hamiltonian in the bulk is given by
\begin{equation}
H=-J\sum\limits_{\langle ij\rangle}P_{ij} - J'\sum\limits_{\langle ij\rangle'}P_{ij},
\end{equation}
where $P_{ij}=1/4-\mbf{S}_{i}\cdot\mbf{S}_{j}$ is the projection operator to the spin singlet state, while $J$ and $J'$ are the AF exchange interaction strengths on the weak (black) bonds and the strong (blue) bonds, respectively.

The ground state in the bulk has a long-range AF order for $J'/J \simeq 1$, and is strongly dimerized and disordered for $J'/J \gg 1$. A continuous quantum phase transition takes place at $J'/J = 1.90951(1)$ and belongs to the 3D O(3) universality class~\cite{Matsumoto2001a, Ma2018}. We set $J=1$ and keep $J'/J$ at the critical value throughout this work.

We focus on the open boundary with a dangling spin chain weakly coupled to the critical bulk state, which is illustrated in Fig.~\ref{fig:model}(a). This boundary configuration shows nonordinary critical behavior, which is distinct from the boundary without the dangling spin chain~\cite{Ding2018, Weber2018}. We introduce the following multispin $Q$-term along the spin chain to explore its boundary phase diagram,
\begin{equation}
H_{Q}=-Q\sum_{l}P_{l,l+1}P_{l+2,l+3}P_{l+4,l+5}, \label{eq:qterm}
\end{equation}
which is illustrated in Fig.~\ref{fig:model}(b). A large $Q$-term is expected to induce a VBS order along the spin chain, which spontaneously breaks the lattice translation symmetry. The $Q$-term was initially introduced in the 2D square lattice~\cite{Sandvik2007, Lou2009a} to drive an AF-VBS transition, which was proposed to be a deconfined QCP~\cite{Senthil2004a, Senthil2004b}.

In our simulations, the lattice has open boundaries with dangling spin chains in one direction and periodic boundary condition in the other direction. The lattice has $L\times L$ sites, with the linear system size $L$ up to $96$. The projective quantum Monte Carlo (QMC) algorithm in the valence bond basis~\cite{Sandvik2005, Sandvik2010a} is adopted to calculate the ground state properties. More than $10^{7}$ Monte Carlo sweeps are performed for each value of the tuning parameter $Q$. The physical quantities and the data processing procedure will be introduced in the following sections.

\begin{figure*}[tb]
\centering
\includegraphics[width=\textwidth]{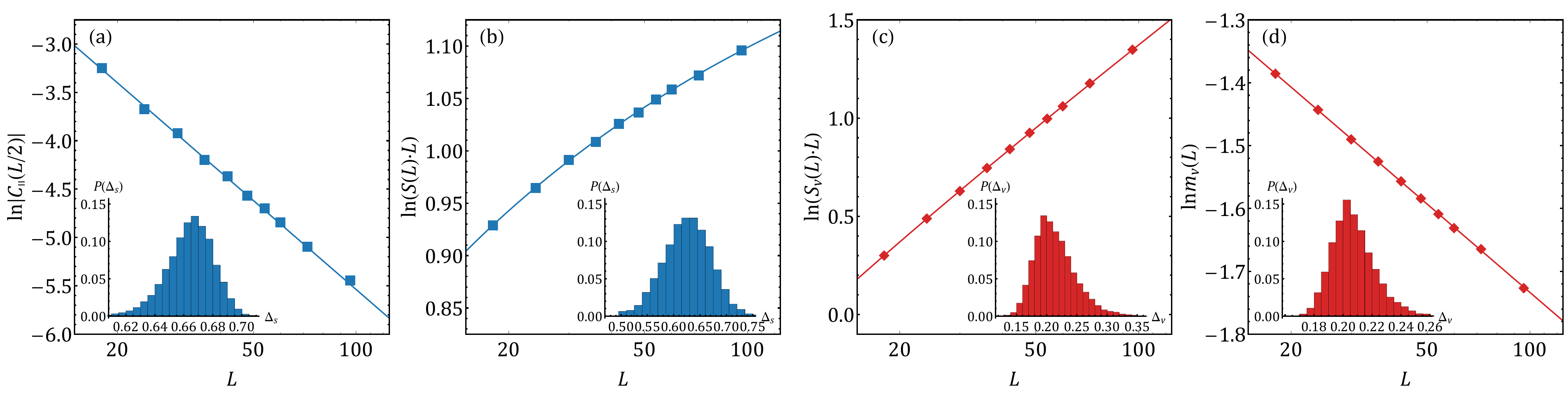}
\caption{Scaling dimensions of boundary order parameters at $Q_{c}$. (a) Staggered spin correlation function, (b) static AF structure factor, (c) VBS structure factor, and (d) VBS order parameter versus the lattice size $L$ in log-log scale. The solid curves are fitting functions according to the corresponding scaling relations. Insets: Histograms of the estimated values of scaling dimensions from the resampling procedure.}
\label{fig:cor}
\end{figure*}

{\it Boundary critical point.---}
Let us first locate the boundary critical point induced by the $Q$-term by examining the Binder ratio of the VBS order parameter, $R_{v}=\langle v^{2}\rangle/\langle|v|\rangle^{2}$, where
\begin{equation}
v=L^{-1}\sum_{l}(-1)^{l}\mbf{S}_{l}\cdot\mbf{S}_{l+1}
\end{equation}
is defined along the dangling spin chain. The Binder ratio is scale invariant at a critical point, thus the curves $R_{v}(Q,L)$ for different lattice sizes should intersect at the critical point up to slight deviation due to irrelevant corrections. From the numerical results plotted in Fig.~\ref{fig:binder}(a), the intersection is evident and indicates a boundary transition. The crossing point between the curves of $R_{v}(Q,L)$ and $R_{v}(Q,2L)$, denoted by $Q_{c}(L)$, gives an estimate of the critical point, which is extracted by interpolating both curves with polynomials and solving the crossing point. $Q_{c}(L)$ approaches the critical point $Q_{c}$ in the thermodynamic limit by
\begin{equation}
Q_{c}(L)=Q_{c}+aL^{-\omega},
\label{eq:qc}
\end{equation}
where $\omega>0$ is the exponent associated with irrelevant corrections. Fitting Eq.~(\ref{eq:qc}) to the crossing points $Q_{c}(L)$ shown in Fig.~\ref{fig:binder}(b) yields $Q_{c}=0.310(11)$. Here, the error bar is a composition of the uncertainty in the nonlinear least-square fitting and that due to the statistical fluctuations of the raw data. The statistical uncertainty is estimated with the following resampling procedure: Assuming that each $R_{v}(Q,L)$ obeys a Gaussian distribution with the standard deviation given by its error bar in the QMC results, many ($\sim 10^{4}$) sets of artificial $R_{v}(Q,L)$ data are generated from these Gaussian distributions and yield the corresponding estimated $Q_{c}$ values with the crossing point analysis, whose histogram is shown in the inset of Fig.~\ref{fig:binder}(b). The standard deviation of the histogram quantifies the statistical uncertainty of $Q_{c}$. This resampling procedure is also used to estimate the error bars of the following critical exponents.

The slope of the Binder ratio at $Q_{c}$, $s_{v}(Q_{c},L)=\dd R_{v}(Q_{c},L)/\dd Q$, grows with the lattice size by a power law,
\begin{equation}
s_{v}(Q_{c},L)\propto L^{y_{s}},
\label{eq:ys}
\end{equation}
where $y_{s}$ is the critical exponent associated with the relevant operator that drives the boundary transition. The slopes shown in Fig.~\ref{fig:binder}(c) are extracted by interpolating the data $R_{v}(Q,L)$ around $Q_{c}$ with polynomials and taking the derivatives. The power-law fitting according to Eq.~(\ref{eq:ys}) yields $y_{s}=0.81(4)$.

The scaling dimension of the AF order parameter at the critical point is defined by the power-law decay of the staggered spin correlations function,
\begin{equation}
C_{\parallel}(i-j)=\langle \mbf{S}_{i}\cdot \mbf{S}_{j}\rangle \propto (-1)^{i-j}|i-j|^{-2\Delta_{s}}. \label{eq:cs}
\end{equation}
The staggered spin correlation function $(-1)^{L/2}C_{\parallel}(L/2)$ versus $L$ is plotted in Fig.~\ref{fig:cor}(a). The power-law fitting yields the scaling dimension $\Delta_{s}=0.664(16)$. The static AF structure factor is obtained by summing $(-1)^{r}C_{\parallel}(r)$ over the distance $r$ along the chain, and thus scales as
\begin{equation}
S(L)\cdot L=L^{-1}\sum_{ij}(-1)^{i-j}C_{\parallel}(i-j) = c+aL^{1-2\Delta_{s}}, \label{eq:sl}
\end{equation}
where the constant term comes from the nonuniversal short-range contribution. Fitting Eq.~(\ref{eq:sl}) to the data shown in Fig.~\ref{fig:cor}(b) yields $\Delta_{s}=0.621(46)$, which is consistent with the previous result within error bars. Combining these results, our final estimate is $\Delta_{s}=0.660(15)$.

The boundary VBS structure factor has the similar scaling form as Eq.~(\ref{eq:sl}) of the AF structure factor,
\begin{equation}
S_{v}(L)\cdot L=\langle v^{2}\rangle \cdot L = c'+a'L^{1-2\Delta_{v}},
\end{equation}
where $\Delta_{v}$ is the scaling dimension of the VBS order parameter. The VBS order parameter scales as
\begin{equation}
m_{v}=\langle|v|\rangle \propto L^{-\Delta_{v}}.
\end{equation}
Fitting these relations to the data plotted in Fig.~\ref{fig:cor}(c) and (d), we find the scaling dimension $\Delta_{v}=0.213(46)$ and $\Delta_{v}=0.203(15)$, respectively, which are also consistent with each other within error bars. Combining these results gives our final estimate $\Delta_{v}=0.204(14)$.

These scaling dimensions at the boundary critical point were calculated perturbatively in Ref.~\cite{Jian2021}, and the results are
\begin{equation}
\Delta_{s}=1/2+\epsilon_{n},\quad \Delta_{v}=1/2-3\epsilon_{n},
\end{equation}
where $\epsilon_{n}$ is the anomalous scaling dimension of the boundary AF order parameter in the ordinary class, $\epsilon_{n}=3/2-\Delta_{s}^{\mr{ord}}=0.306$~\cite{Deng2005, ParisenToldin2023}. Although these results are not quantitatively accurate, they predict the correct signs of $\Delta_{s,v}-1/2$, i.e., the deviation of the scaling dimensions from those of the isolated Heisenberg chain. This suggests that the perturbative analysis is qualitatively correct, and higher-order perturbation may provide more accurate prediction of the scaling dimensions.

\begin{figure*}[tb]
\centering
\includegraphics[width=\textwidth]{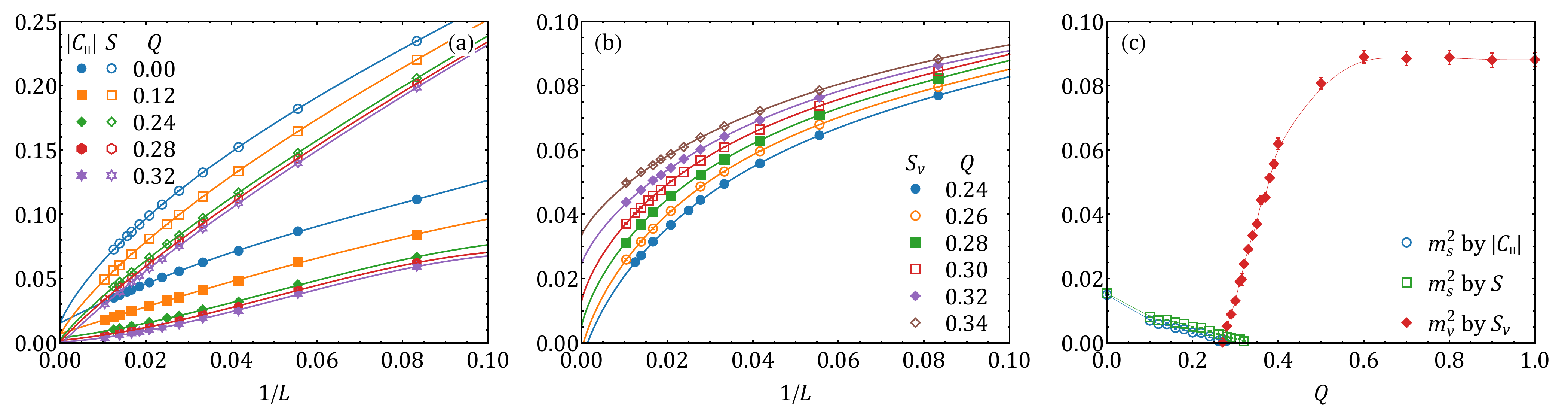}
\caption{Order parameters on the boundary. (a) The staggered spin correlation function $(-1)^{L/2}C_{\parallel}(L/2)$ and the static AF structure factor $S(L)$ fitted with Eqs.~(\ref{eq:ms2c}) and (\ref{eq:sml}) to extrapolate the squared AF order parameter $m_{s}^{2}$. (b) The VBS structure factor $S_{v}(L)$ fitted with Eq.~(\ref{eq:svl}) to extrapolate $m_{v}^{2}$. (c) The squared order parameters $m_{s}^{2}$ and $m_{v}^{2}$ versus the tuning parameter $Q$. $m_{s}^{2}$ vanishes as $Q\to Q_{c}^{-}$, while $m_{v}^{2}$ vanishes as $Q\to Q_{c}^{+}$, indicating a continuous AF-VBS transition on the boundary.}
\label{fig:order}
\end{figure*}

{\it Ordered phases.---}
We then focus on the boundary phases away from the critical point. For $Q<Q_{c}$, the AF correlation on the boundary is strongly enhanced. According to the theoretical analysis, the spin correlation function may either saturate to a long-range AF order~\cite{Jian2021, Song2025}, or decay logarithmically to zero~\cite{Metlitski2022}, hence we examine both possibilities below. If the boundary has a long-range AF order, the staggered spin correlation function can be expanded as a polynomial of $r^{-1}$ as $r\to \infty$,
\begin{equation}
(-1)^{r}C_{\parallel}(r)=m_{s}^{2}+a_{1}r^{-1}+a_{2}r^{-2}+\ldots, \label{eq:ms2c}
\end{equation}
then the static AF structure factor should scale as
\begin{equation}
S(L)=m_{s}^{2}+a_{0}'L^{-1}\ln L+a_{1}'L^{-1}+a_{2}'L^{-2}+\ldots, \label{eq:sml}
\end{equation}
where the $L^{-1}\ln L$ term comes from summing over the $r^{-1}$ term in Eq.~(\ref{eq:ms2c}). Fitting these relations to the numerical results shown in Fig.~\ref{fig:order}(a), we find nonzero squared AF order parameter $m_{s}^{2}$ for $Q<Q_{c}$, which vanishes as $Q\to Q_{c}^{-}$ [see Fig.~\ref{fig:order}(c)]. The values of $m_{s}^{2}$ extracted from $(-1)^{L/2} C_{\parallel}(L/2)$ and $S(L)$ are consistent with each other within one or two error bars. On the other hand, the results of $(-1)^{L/2} C_{\parallel}(L/2)$ and $S(L)$ cannot be adequately described by the extraordinary-log scaling form (see the Appendix for the trial fitting details). Therefore, we conclude that the boundary has a long-range AF order for $Q<Q_{c}$, which is so weak that it was overlooked in previous works~\cite{Ding2018, Weber2018}. This AF order does not violate the Mermin-Wagner theorem~\cite{Mermin1966}, because the coupling with the critical bulk state induces qLR effective interactions with the strength decaying by power law with the spacetime distance~\cite{Jian2021}.

\begin{figure}[tb]
\centering
\includegraphics[width=\columnwidth]{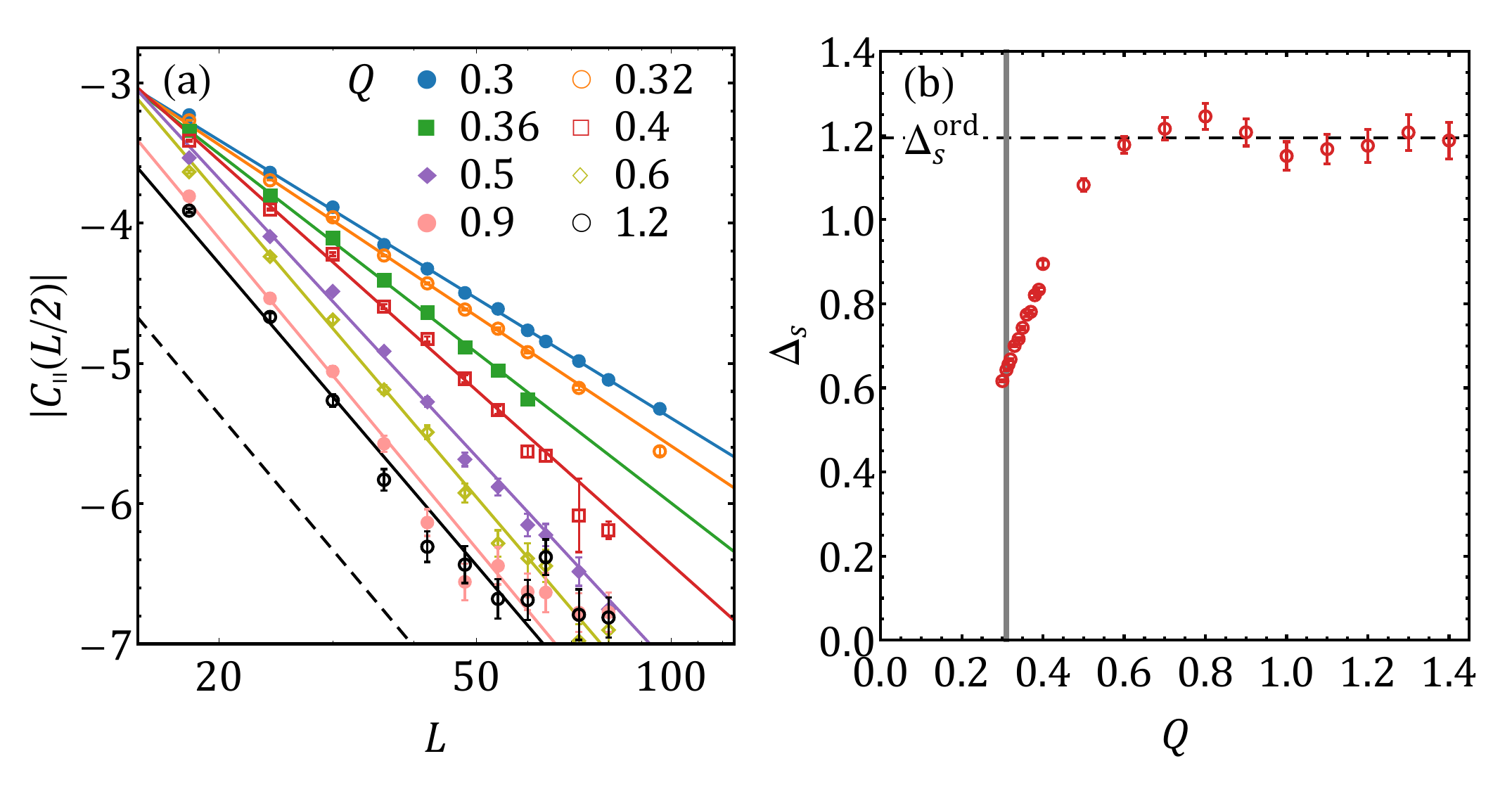}
\caption{Boundary critical behavior in the VBS phase. (a) Staggered spin correlation function $(-1)^{L/2}C_{\parallel}(L/2)$ versus $L$ in the log-log scale. The solid lines are power-law fitting to extract the scaling dimension $\Delta_{s}$. The power-law decay of the ordinary class is drawn as the dashed line for comparison. (b) The extracted scaling dimension $\Delta_{s}$ versus the tuning parameter $Q$. The scaling dimension $\Delta_{s}^{\mr{ord}}=1.194$ in the ordinary class~\cite{Deng2005, ParisenToldin2023} is drawn as the dashed line for comparison.}
\label{fig:ordinary}
\end{figure}

For $Q>Q_{c}$, a long-range VBS order is induced on the boundary. Similar to the expansion in Eq.~(\ref{eq:sml}), we fit the VBS structure factor $S_{v}(L)$ shown in Fig.~\ref{fig:order}(b) with the following form,
\begin{equation}
S_{v}(L)=m_{v}^{2}+b_{0}'L^{-1}\ln L+b_{1}'L^{-1}+b_{2}'L^{-2}+\ldots, \label{eq:svl}
\end{equation}
and find a nonzero squared VBS order parameter $m_{v}^{2}$ for $Q>Q_{c}$, which vanishes as $Q\to Q_{c}^{+}$ [see Fig.~\ref{fig:order}(c)]. Therefore, the critical point $Q_{c}$ is a boundary transition between the AF and the VBS long-range orders, which is reminiscent of the deconfined QCP in (2+1)D frustrated quantum antiferromagnets~\cite{Senthil2004a, Senthil2004b}. While the low-energy excitations in the AF phase are spin waves, the elementary excitations in the VBS phase are deconfined spin-$1/2$ spinons on the VBS domain walls~\cite{Tang2011a, Tang2015}. This justifies our notion of a deconfined boundary transition of the quantum critical Heisenberg model.

We then examine the boundary critical behavior in the VBS phase. The staggered spin correlation functions $(-1)^{L/2}C_{\parallel}(L/2)$ versus $L$ for various $Q \gtrsim Q_{c}$ are plotted in Fig.~\ref{fig:ordinary}(a), and fitted with the power law in Eq.~(\ref{eq:cs}). The extracted $\Delta_{s}$ gradually increases from its value at $Q_{c}$ and saturates deep in the VBS phase to the scaling dimension of the ordinary class, $\Delta_{s}^{\mr{ord}}=1.194$~\cite{Deng2005, ParisenToldin2023}. The crossover near $Q_{c}$ can be attributed to the finite-size effect. Therefore, the boundary in the VBS phase exhibits ordinary critical behavior, thus confirming the theoretical prediction in Ref.~\cite{Jian2021, Song2025}.

\begin{ruledtabular}
\begin{table}[!tb]
\caption{Long-range behavior of correlation functions in 2D NL$\sigma$M with/without topological $\theta$-term and/or qLR interactions.}
\label{tab:nlsm}
\begin{tabular}{c| c c}
2D NL$\sigma$M				&	without $\theta$-term	&	with $\theta$-term, $\theta=\pi$	\\
\hline
without qLR coupling		&	exp. decay				&	power-law decay						\\
with qLR coupling			&	log. decay				&	long-range order
\end{tabular}
\end{table}
\end{ruledtabular}

{\it Conclusion and discussions.---}
We have numerically identified a deconfined AF-VBS transition on the boundary of a (2+1)D quantum critical Heisenberg model. A long-range AF order is observed on the dangling spin chain weakly coupled to the critical bulk state. It is gradually suppressed by the multispin $Q$-term along the chain, giving way to a VBS order for $Q>Q_{c}$ with ordinary boundary critical behavior. This phase diagram is consistent with the theoretical prediction based on the bosonized spin chain coupled with the critical bulk state~\cite{Jian2021, Song2025}.

The underlying physics is governed by the interplay between the topological $\theta$-term in the 2D O(3) NL$\sigma$M~\cite{Haldane1985, Haldane1988} and the qLR effective interactions mediated by the critical bulk state. While the $\theta$-term alone partially suppresses fluctuations via destructive interference of different topological sectors and leads to the power-law decaying spin correlation function of the AF Heisenberg chain~\cite{Witten1984, Affleck1985b, Affleck1985a}, and qLR interactions alone lead to the extraordinary-log behavior~\cite{Metlitski2022}, their synergy suppresses both smooth fluctuations and spacetime monopole events, thereby stabilizing the long-range AF order (Table~\ref{tab:nlsm}). Our findings demonstrate that such interplay enables rich boundary critical behavior beyond the Landau paradigm, and suggest exploring the boundary phase diagrams in bulk topological states with gapless boundaries.

\begin{acknowledgments}
C.D. is supported by the National Natural Science Foundation of China (No.~11975024). L.Z. is supported by the National Natural Science Foundation of China (No.~12174387), the Chinese Academy of Sciences (Nos.~XDB1700000 and JZHKYPT-2021-08), and the Innovative Program for Quantum Science and Technology (No.~2021ZD0302600).
\end{acknowledgments}

\bibliography{C:/Documents/ResearchNotes/BibTex/library}

\begin{thebibliography}{43}%
\makeatletter
\providecommand \@ifxundefined [1]{%
 \@ifx{#1\undefined}
}%
\providecommand \@ifnum [1]{%
 \ifnum #1\expandafter \@firstoftwo
 \else \expandafter \@secondoftwo
 \fi
}%
\providecommand \@ifx [1]{%
 \ifx #1\expandafter \@firstoftwo
 \else \expandafter \@secondoftwo
 \fi
}%
\providecommand \natexlab [1]{#1}%
\providecommand \enquote  [1]{``#1''}%
\providecommand \bibnamefont  [1]{#1}%
\providecommand \bibfnamefont [1]{#1}%
\providecommand \citenamefont [1]{#1}%
\providecommand \href@noop [0]{\@secondoftwo}%
\providecommand \href [0]{\begingroup \@sanitize@url \@href}%
\providecommand \@href[1]{\@@startlink{#1}\@@href}%
\providecommand \@@href[1]{\endgroup#1\@@endlink}%
\providecommand \@sanitize@url [0]{\catcode `\\12\catcode `\$12\catcode
  `\&12\catcode `\#12\catcode `\^12\catcode `\_12\catcode `\%12\relax}%
\providecommand \@@startlink[1]{}%
\providecommand \@@endlink[0]{}%
\providecommand \url  [0]{\begingroup\@sanitize@url \@url }%
\providecommand \@url [1]{\endgroup\@href {#1}{\urlprefix }}%
\providecommand \urlprefix  [0]{URL }%
\providecommand \Eprint [0]{\href }%
\providecommand \doibase [0]{https://doi.org/}%
\providecommand \selectlanguage [0]{\@gobble}%
\providecommand \bibinfo  [0]{\@secondoftwo}%
\providecommand \bibfield  [0]{\@secondoftwo}%
\providecommand \translation [1]{[#1]}%
\providecommand \BibitemOpen [0]{}%
\providecommand \bibitemStop [0]{}%
\providecommand \bibitemNoStop [0]{.\EOS\space}%
\providecommand \EOS [0]{\spacefactor3000\relax}%
\providecommand \BibitemShut  [1]{\csname bibitem#1\endcsname}%
\let\auto@bib@innerbib\@empty
\bibitem [{\citenamefont {Binder}(1983)}]{Binder1983phase}%
  \BibitemOpen
  \bibfield  {author} {\bibinfo {author} {\bibfnamefont {K.}~\bibnamefont
  {Binder}},\ }\bibfield  {title} {\bibinfo {title} {{Critical behaviour at
  surfaces}},\ }in\ \href@noop {} {\emph {\bibinfo {booktitle} {Phase
  Transitions Crit. Phenom.}}},\ Vol.~\bibinfo {volume} {8},\ \bibinfo {editor}
  {edited by\ \bibinfo {editor} {\bibfnamefont {C.}~\bibnamefont {Domb}}\ and\
  \bibinfo {editor} {\bibfnamefont {J.~L.}\ \bibnamefont {Lebowitz}}}\
  (\bibinfo  {publisher} {Academic Press},\ \bibinfo {address} {London,
  England},\ \bibinfo {year} {1983})\BibitemShut {NoStop}%
\bibitem [{\citenamefont {Zhang}\ and\ \citenamefont {Wang}(2017)}]{Zhang2017}%
  \BibitemOpen
  \bibfield  {author} {\bibinfo {author} {\bibfnamefont {L.}~\bibnamefont
  {Zhang}}\ and\ \bibinfo {author} {\bibfnamefont {F.}~\bibnamefont {Wang}},\
  }\bibfield  {title} {\bibinfo {title} {{Unconventional surface critical
  behavior induced by a quantum phase transition from the two-dimensional
  Affleck-Kennedy-Lieb-Tasaki phase to a N{\'{e}}el-ordered phase}},\ }\href
  {https://doi.org/10.1103/PhysRevLett.118.087201} {\bibfield  {journal}
  {\bibinfo  {journal} {Phys. Rev. Lett.}\ }\textbf {\bibinfo {volume} {118}},\
  \bibinfo {pages} {087201} (\bibinfo {year} {2017})}\BibitemShut {NoStop}%
\bibitem [{\citenamefont {Ding}\ \emph {et~al.}(2018)\citenamefont {Ding},
  \citenamefont {Zhang},\ and\ \citenamefont {Guo}}]{Ding2018}%
  \BibitemOpen
  \bibfield  {author} {\bibinfo {author} {\bibfnamefont {C.}~\bibnamefont
  {Ding}}, \bibinfo {author} {\bibfnamefont {L.}~\bibnamefont {Zhang}},\ and\
  \bibinfo {author} {\bibfnamefont {W.}~\bibnamefont {Guo}},\ }\bibfield
  {title} {\bibinfo {title} {{Engineering surface critical behavior of
  (2+1)-dimensional O(3) quantum critical points}},\ }\href
  {https://doi.org/10.1103/PhysRevLett.120.235701} {\bibfield  {journal}
  {\bibinfo  {journal} {Phys. Rev. Lett.}\ }\textbf {\bibinfo {volume} {120}},\
  \bibinfo {pages} {235701} (\bibinfo {year} {2018})}\BibitemShut {NoStop}%
\bibitem [{\citenamefont {Weber}\ \emph {et~al.}(2018)\citenamefont {Weber},
  \citenamefont {{Parisen Toldin}},\ and\ \citenamefont {Wessel}}]{Weber2018}%
  \BibitemOpen
  \bibfield  {author} {\bibinfo {author} {\bibfnamefont {L.}~\bibnamefont
  {Weber}}, \bibinfo {author} {\bibfnamefont {F.}~\bibnamefont {{Parisen
  Toldin}}},\ and\ \bibinfo {author} {\bibfnamefont {S.}~\bibnamefont
  {Wessel}},\ }\bibfield  {title} {\bibinfo {title} {{Nonordinary edge
  criticality of two-dimensional quantum critical magnets}},\ }\href
  {https://doi.org/10.1103/PhysRevB.98.140403} {\bibfield  {journal} {\bibinfo
  {journal} {Phys. Rev. B}\ }\textbf {\bibinfo {volume} {98}},\ \bibinfo
  {pages} {140403} (\bibinfo {year} {2018})}\BibitemShut {NoStop}%
\bibitem [{\citenamefont {Weber}\ and\ \citenamefont
  {Wessel}(2019)}]{Weber2019a}%
  \BibitemOpen
  \bibfield  {author} {\bibinfo {author} {\bibfnamefont {L.}~\bibnamefont
  {Weber}}\ and\ \bibinfo {author} {\bibfnamefont {S.}~\bibnamefont {Wessel}},\
  }\bibfield  {title} {\bibinfo {title} {{Nonordinary criticality at the edges
  of planar spin-1 Heisenberg antiferromagnets}},\ }\href
  {https://doi.org/10.1103/PhysRevB.100.054437} {\bibfield  {journal} {\bibinfo
   {journal} {Phys. Rev. B}\ }\textbf {\bibinfo {volume} {100}},\ \bibinfo
  {pages} {054437} (\bibinfo {year} {2019})}\BibitemShut {NoStop}%
\bibitem [{\citenamefont {Zhu}\ \emph {et~al.}(2021)\citenamefont {Zhu},
  \citenamefont {Ding}, \citenamefont {Zhang},\ and\ \citenamefont
  {Guo}}]{Zhu2021}%
  \BibitemOpen
  \bibfield  {author} {\bibinfo {author} {\bibfnamefont {W.}~\bibnamefont
  {Zhu}}, \bibinfo {author} {\bibfnamefont {C.}~\bibnamefont {Ding}}, \bibinfo
  {author} {\bibfnamefont {L.}~\bibnamefont {Zhang}},\ and\ \bibinfo {author}
  {\bibfnamefont {W.}~\bibnamefont {Guo}},\ }\bibfield  {title} {\bibinfo
  {title} {{Surface critical behavior of coupled Haldane chains}},\ }\href
  {https://doi.org/10.1103/PhysRevB.103.024412} {\bibfield  {journal} {\bibinfo
   {journal} {Phys. Rev. B}\ }\textbf {\bibinfo {volume} {103}},\ \bibinfo
  {pages} {024412} (\bibinfo {year} {2021})}\BibitemShut {NoStop}%
\bibitem [{\citenamefont {Weber}\ and\ \citenamefont
  {Wessel}(2021)}]{Weber2021}%
  \BibitemOpen
  \bibfield  {author} {\bibinfo {author} {\bibfnamefont {L.}~\bibnamefont
  {Weber}}\ and\ \bibinfo {author} {\bibfnamefont {S.}~\bibnamefont {Wessel}},\
  }\bibfield  {title} {\bibinfo {title} {{Spin versus bond correlations along
  dangling edges of quantum critical magnets}},\ }\href
  {https://doi.org/10.1103/PhysRevB.103.L020406} {\bibfield  {journal}
  {\bibinfo  {journal} {Phys. Rev. B}\ }\textbf {\bibinfo {volume} {103}},\
  \bibinfo {pages} {L020406} (\bibinfo {year} {2021})}\BibitemShut {NoStop}%
\bibitem [{\citenamefont {Zhu}\ \emph {et~al.}()\citenamefont {Zhu},
  \citenamefont {Ding}, \citenamefont {Zhang},\ and\ \citenamefont
  {Guo}}]{Zhu2021b}%
  \BibitemOpen
  \bibfield  {author} {\bibinfo {author} {\bibfnamefont {W.}~\bibnamefont
  {Zhu}}, \bibinfo {author} {\bibfnamefont {C.}~\bibnamefont {Ding}}, \bibinfo
  {author} {\bibfnamefont {L.}~\bibnamefont {Zhang}},\ and\ \bibinfo {author}
  {\bibfnamefont {W.}~\bibnamefont {Guo}},\ }\bibfield  {title} {\bibinfo
  {title} {{Exotic surface behaviors induced by geometrical settings of
  two-dimensional dimerized quantum XXZ model}},\ }\href
  {http://arxiv.org/abs/2111.12336} {\ }\Eprint
  {https://arxiv.org/abs/2111.12336} {arXiv:2111.12336} \BibitemShut {NoStop}%
\bibitem [{\citenamefont {Wang}\ \emph {et~al.}(2022)\citenamefont {Wang},
  \citenamefont {Zhang},\ and\ \citenamefont {Guo}}]{Wang2022}%
  \BibitemOpen
  \bibfield  {author} {\bibinfo {author} {\bibfnamefont {Z.}~\bibnamefont
  {Wang}}, \bibinfo {author} {\bibfnamefont {F.}~\bibnamefont {Zhang}},\ and\
  \bibinfo {author} {\bibfnamefont {W.}~\bibnamefont {Guo}},\ }\bibfield
  {title} {\bibinfo {title} {{Bulk and surface critical behavior of a quantum
  Heisenberg antiferromagnet on two-dimensional coupled diagonal ladders}},\
  }\href {https://doi.org/10.1103/PhysRevB.106.134407} {\bibfield  {journal}
  {\bibinfo  {journal} {Phys. Rev. B}\ }\textbf {\bibinfo {volume} {106}},\
  \bibinfo {pages} {134407} (\bibinfo {year} {2022})}\BibitemShut {NoStop}%
\bibitem [{\citenamefont {Wang}\ \emph {et~al.}(2023)\citenamefont {Wang},
  \citenamefont {Zhang},\ and\ \citenamefont {Guo}}]{Wang2023d}%
  \BibitemOpen
  \bibfield  {author} {\bibinfo {author} {\bibfnamefont {Z.}~\bibnamefont
  {Wang}}, \bibinfo {author} {\bibfnamefont {F.}~\bibnamefont {Zhang}},\ and\
  \bibinfo {author} {\bibfnamefont {W.}~\bibnamefont {Guo}},\ }\bibfield
  {title} {\bibinfo {title} {{Extraordinary surface critical behavior induced
  by the symmetry-protected topological states of a two-dimensional quantum
  magnet}},\ }\href {https://doi.org/10.1103/PhysRevB.108.014409} {\bibfield
  {journal} {\bibinfo  {journal} {Phys. Rev. B}\ }\textbf {\bibinfo {volume}
  {108}},\ \bibinfo {pages} {014409} (\bibinfo {year} {2023})}\BibitemShut
  {NoStop}%
\bibitem [{\citenamefont {Ding}\ \emph {et~al.}(2023)\citenamefont {Ding},
  \citenamefont {Zhu}, \citenamefont {Guo},\ and\ \citenamefont
  {Zhang}}]{Ding2023}%
  \BibitemOpen
  \bibfield  {author} {\bibinfo {author} {\bibfnamefont {C.}~\bibnamefont
  {Ding}}, \bibinfo {author} {\bibfnamefont {W.}~\bibnamefont {Zhu}}, \bibinfo
  {author} {\bibfnamefont {W.}~\bibnamefont {Guo}},\ and\ \bibinfo {author}
  {\bibfnamefont {L.}~\bibnamefont {Zhang}},\ }\bibfield  {title} {\bibinfo
  {title} {{Special transition and extraordinary phase on the surface of a
  two-dimensional quantum Heisenberg antiferromagnet}},\ }\href
  {https://doi.org/10.21468/SciPostPhys.15.1.012} {\bibfield  {journal}
  {\bibinfo  {journal} {SciPost Phys.}\ }\textbf {\bibinfo {volume} {15}},\
  \bibinfo {pages} {012} (\bibinfo {year} {2023})}\BibitemShut {NoStop}%
\bibitem [{\citenamefont {{Parisen Toldin}}(2023)}]{ParisenToldin2023}%
  \BibitemOpen
  \bibfield  {author} {\bibinfo {author} {\bibfnamefont {F.}~\bibnamefont
  {{Parisen Toldin}}},\ }\bibfield  {title} {\bibinfo {title} {{The ordinary
  surface universality class of the three-dimensional O(N) model}},\ }\href
  {https://doi.org/10.1103/PhysRevB.108.L020404} {\bibfield  {journal}
  {\bibinfo  {journal} {Phys. Rev. B}\ }\textbf {\bibinfo {volume} {108}},\
  \bibinfo {pages} {L020404} (\bibinfo {year} {2023})}\BibitemShut {NoStop}%
\bibitem [{\citenamefont {{Parisen Toldin}}(2021)}]{ParisenToldin2021}%
  \BibitemOpen
  \bibfield  {author} {\bibinfo {author} {\bibfnamefont {F.}~\bibnamefont
  {{Parisen Toldin}}},\ }\bibfield  {title} {\bibinfo {title} {{Boundary
  critical behavior of the three-dimensional Heisenberg universality class}},\
  }\href {https://doi.org/10.1103/PhysRevLett.126.135701} {\bibfield  {journal}
  {\bibinfo  {journal} {Phys. Rev. Lett.}\ }\textbf {\bibinfo {volume} {126}},\
  \bibinfo {pages} {135701} (\bibinfo {year} {2021})}\BibitemShut {NoStop}%
\bibitem [{\citenamefont {Hu}\ \emph {et~al.}(2021)\citenamefont {Hu},
  \citenamefont {Deng},\ and\ \citenamefont {Lv}}]{Hu2021}%
  \BibitemOpen
  \bibfield  {author} {\bibinfo {author} {\bibfnamefont {M.}~\bibnamefont
  {Hu}}, \bibinfo {author} {\bibfnamefont {Y.}~\bibnamefont {Deng}},\ and\
  \bibinfo {author} {\bibfnamefont {J.-P.}\ \bibnamefont {Lv}},\ }\bibfield
  {title} {\bibinfo {title} {{Extraordinary-log surface phase transition in the
  three-dimensional XY model}},\ }\href
  {https://doi.org/10.1103/PhysRevLett.127.120603} {\bibfield  {journal}
  {\bibinfo  {journal} {Phys. Rev. Lett.}\ }\textbf {\bibinfo {volume} {127}},\
  \bibinfo {pages} {120603} (\bibinfo {year} {2021})}\BibitemShut {NoStop}%
\bibitem [{\citenamefont {{Parisen Toldin}}\ and\ \citenamefont
  {Metlitski}(2022)}]{ParisenToldin2022}%
  \BibitemOpen
  \bibfield  {author} {\bibinfo {author} {\bibfnamefont {F.}~\bibnamefont
  {{Parisen Toldin}}}\ and\ \bibinfo {author} {\bibfnamefont {M.~A.}\
  \bibnamefont {Metlitski}},\ }\bibfield  {title} {\bibinfo {title} {{Boundary
  criticality of the 3d O(N) model: from normal to extraordinary}},\ }\href
  {https://doi.org/10.1103/PhysRevLett.128.215701} {\bibfield  {journal}
  {\bibinfo  {journal} {Phys. Rev. Lett.}\ }\textbf {\bibinfo {volume} {128}},\
  \bibinfo {pages} {215701} (\bibinfo {year} {2022})}\BibitemShut {NoStop}%
\bibitem [{\citenamefont {Zhang}\ \emph {et~al.}(2022)\citenamefont {Zhang},
  \citenamefont {Ding}, \citenamefont {Deng},\ and\ \citenamefont
  {Zhang}}]{Zhang2022}%
  \BibitemOpen
  \bibfield  {author} {\bibinfo {author} {\bibfnamefont {L.-R.~L.}\
  \bibnamefont {Zhang}}, \bibinfo {author} {\bibfnamefont {C.}~\bibnamefont
  {Ding}}, \bibinfo {author} {\bibfnamefont {Y.}~\bibnamefont {Deng}},\ and\
  \bibinfo {author} {\bibfnamefont {L.-R.~L.}\ \bibnamefont {Zhang}},\
  }\bibfield  {title} {\bibinfo {title} {{Surface criticality of the
  antiferromagnetic Potts model}},\ }\href
  {https://doi.org/10.1103/PhysRevB.105.224415} {\bibfield  {journal} {\bibinfo
   {journal} {Phys. Rev. B}\ }\textbf {\bibinfo {volume} {105}},\ \bibinfo
  {pages} {224415} (\bibinfo {year} {2022})}\BibitemShut {NoStop}%
\bibitem [{\citenamefont {Zhang}\ \emph {et~al.}(2023)\citenamefont {Zhang},
  \citenamefont {Ding}, \citenamefont {Zhang},\ and\ \citenamefont
  {Zhang}}]{Zhang2023a}%
  \BibitemOpen
  \bibfield  {author} {\bibinfo {author} {\bibfnamefont {L.-R.~L.}\
  \bibnamefont {Zhang}}, \bibinfo {author} {\bibfnamefont {C.}~\bibnamefont
  {Ding}}, \bibinfo {author} {\bibfnamefont {W.}~\bibnamefont {Zhang}},\ and\
  \bibinfo {author} {\bibfnamefont {L.-R.~L.}\ \bibnamefont {Zhang}},\
  }\bibfield  {title} {\bibinfo {title} {{Sublattice extraordinary-log phase
  and special points of the antiferromagnetic Potts model}},\ }\href
  {https://doi.org/10.1103/PhysRevB.108.024402} {\bibfield  {journal} {\bibinfo
   {journal} {Phys. Rev. B}\ }\textbf {\bibinfo {volume} {108}},\ \bibinfo
  {pages} {024402} (\bibinfo {year} {2023})},\ \Eprint
  {https://arxiv.org/abs/2301.08926} {arXiv:2301.08926} \BibitemShut {NoStop}%
\bibitem [{\citenamefont {Zou}\ \emph {et~al.}(2022)\citenamefont {Zou},
  \citenamefont {Liu},\ and\ \citenamefont {Guo}}]{Zou2022}%
  \BibitemOpen
  \bibfield  {author} {\bibinfo {author} {\bibfnamefont {X.}~\bibnamefont
  {Zou}}, \bibinfo {author} {\bibfnamefont {S.}~\bibnamefont {Liu}},\ and\
  \bibinfo {author} {\bibfnamefont {W.}~\bibnamefont {Guo}},\ }\bibfield
  {title} {\bibinfo {title} {{Surface critical properties of the
  three-dimensional clock model}},\ }\href
  {https://doi.org/10.1103/PhysRevB.106.064420} {\bibfield  {journal} {\bibinfo
   {journal} {Phys. Rev. B}\ }\textbf {\bibinfo {volume} {106}},\ \bibinfo
  {pages} {064420} (\bibinfo {year} {2022})}\BibitemShut {NoStop}%
\bibitem [{\citenamefont {Sun}\ and\ \citenamefont {Lv}(2022)}]{Sun2022}%
  \BibitemOpen
  \bibfield  {author} {\bibinfo {author} {\bibfnamefont {Y.}~\bibnamefont
  {Sun}}\ and\ \bibinfo {author} {\bibfnamefont {J.-P.}\ \bibnamefont {Lv}},\
  }\bibfield  {title} {\bibinfo {title} {{Quantum extraordinary-log
  universality of boundary critical behavior}},\ }\href
  {https://doi.org/10.1103/PhysRevB.106.224502} {\bibfield  {journal} {\bibinfo
   {journal} {Phys. Rev. B}\ }\textbf {\bibinfo {volume} {106}},\ \bibinfo
  {pages} {224502} (\bibinfo {year} {2022})}\BibitemShut {NoStop}%
\bibitem [{\citenamefont {Sun}\ \emph {et~al.}(2022)\citenamefont {Sun},
  \citenamefont {Lyu},\ and\ \citenamefont {Lv}}]{Sun2022b}%
  \BibitemOpen
  \bibfield  {author} {\bibinfo {author} {\bibfnamefont {Y.}~\bibnamefont
  {Sun}}, \bibinfo {author} {\bibfnamefont {J.}~\bibnamefont {Lyu}},\ and\
  \bibinfo {author} {\bibfnamefont {J.-P.}\ \bibnamefont {Lv}},\ }\bibfield
  {title} {\bibinfo {title} {{Classical-quantum correspondence of special and
  extraordinary-log criticality: Villain's bridge}},\ }\href
  {https://doi.org/10.1103/PhysRevB.106.174516} {\bibfield  {journal} {\bibinfo
   {journal} {Phys. Rev. B}\ }\textbf {\bibinfo {volume} {106}},\ \bibinfo
  {pages} {174516} (\bibinfo {year} {2022})}\BibitemShut {NoStop}%
\bibitem [{\citenamefont {Jian}\ \emph {et~al.}(2021)\citenamefont {Jian},
  \citenamefont {Xu}, \citenamefont {Wu},\ and\ \citenamefont {Xu}}]{Jian2021}%
  \BibitemOpen
  \bibfield  {author} {\bibinfo {author} {\bibfnamefont {C.-M.}\ \bibnamefont
  {Jian}}, \bibinfo {author} {\bibfnamefont {Y.}~\bibnamefont {Xu}}, \bibinfo
  {author} {\bibfnamefont {X.-C.}\ \bibnamefont {Wu}},\ and\ \bibinfo {author}
  {\bibfnamefont {C.}~\bibnamefont {Xu}},\ }\bibfield  {title} {\bibinfo
  {title} {{Continuous N{\'{e}}el-VBS quantum phase transition in non-local
  one-dimensional systems with SO(3) symmetry}},\ }\href
  {https://doi.org/10.21468/SciPostPhys.10.2.033} {\bibfield  {journal}
  {\bibinfo  {journal} {SciPost Phys.}\ }\textbf {\bibinfo {volume} {10}},\
  \bibinfo {pages} {033} (\bibinfo {year} {2021})}\BibitemShut {NoStop}%
\bibitem [{\citenamefont {Metlitski}(2022)}]{Metlitski2022}%
  \BibitemOpen
  \bibfield  {author} {\bibinfo {author} {\bibfnamefont {M.}~\bibnamefont
  {Metlitski}},\ }\bibfield  {title} {\bibinfo {title} {{Boundary criticality
  of the O(N) model in d=3 critically revisited}},\ }\href
  {https://doi.org/10.21468/SciPostPhys.12.4.131} {\bibfield  {journal}
  {\bibinfo  {journal} {SciPost Phys.}\ }\textbf {\bibinfo {volume} {12}},\
  \bibinfo {pages} {131} (\bibinfo {year} {2022})}\BibitemShut {NoStop}%
\bibitem [{\citenamefont {Padayasi}\ \emph {et~al.}(2022)\citenamefont
  {Padayasi}, \citenamefont {Krishnan}, \citenamefont {Metlitski},
  \citenamefont {Gruzberg},\ and\ \citenamefont {Meineri}}]{Padayasi2022a}%
  \BibitemOpen
  \bibfield  {author} {\bibinfo {author} {\bibfnamefont {J.}~\bibnamefont
  {Padayasi}}, \bibinfo {author} {\bibfnamefont {A.}~\bibnamefont {Krishnan}},
  \bibinfo {author} {\bibfnamefont {M.}~\bibnamefont {Metlitski}}, \bibinfo
  {author} {\bibfnamefont {I.}~\bibnamefont {Gruzberg}},\ and\ \bibinfo
  {author} {\bibfnamefont {M.}~\bibnamefont {Meineri}},\ }\bibfield  {title}
  {\bibinfo {title} {{The extraordinary boundary transition in the 3d O(N)
  model via conformal bootstrap}},\ }\href
  {https://doi.org/10.21468/SciPostPhys.12.6.190} {\bibfield  {journal}
  {\bibinfo  {journal} {SciPost Phys.}\ }\textbf {\bibinfo {volume} {12}},\
  \bibinfo {pages} {190} (\bibinfo {year} {2022})}\BibitemShut {NoStop}%
\bibitem [{\citenamefont {Song}\ and\ \citenamefont {Zhang}(2025)}]{Song2025}%
  \BibitemOpen
  \bibfield  {author} {\bibinfo {author} {\bibfnamefont {H.-H.}\ \bibnamefont
  {Song}}\ and\ \bibinfo {author} {\bibfnamefont {L.}~\bibnamefont {Zhang}},\
  }\bibfield  {title} {\bibinfo {title} {{Boundary phase transitions of
  two-dimensional quantum critical XXZ model}},\ }\href
  {https://doi.org/10.1103/PhysRevB.111.165139} {\bibfield  {journal} {\bibinfo
   {journal} {Phys. Rev. B}\ }\textbf {\bibinfo {volume} {111}},\ \bibinfo
  {pages} {165139} (\bibinfo {year} {2025})}\BibitemShut {NoStop}%
\bibitem [{\citenamefont {Grover}\ and\ \citenamefont
  {Vishwanath}()}]{Grover2012a}%
  \BibitemOpen
  \bibfield  {author} {\bibinfo {author} {\bibfnamefont {T.}~\bibnamefont
  {Grover}}\ and\ \bibinfo {author} {\bibfnamefont {A.}~\bibnamefont
  {Vishwanath}},\ }\bibfield  {title} {\bibinfo {title} {{Quantum criticality
  in topological insulators and superconductors: emergence of strongly coupled
  Majoranas and supersymmetry}},\ }\href {http://arxiv.org/abs/1206.1332} {\
  }\Eprint {https://arxiv.org/abs/1206.1332} {arXiv:1206.1332} \BibitemShut
  {NoStop}%
\bibitem [{\citenamefont {Suzuki}\ and\ \citenamefont
  {Sato}(2012)}]{Suzuki2012b}%
  \BibitemOpen
  \bibfield  {author} {\bibinfo {author} {\bibfnamefont {T.}~\bibnamefont
  {Suzuki}}\ and\ \bibinfo {author} {\bibfnamefont {M.}~\bibnamefont {Sato}},\
  }\bibfield  {title} {\bibinfo {title} {{Gapless edge states and their
  stability in two-dimensional quantum magnets}},\ }\href
  {https://doi.org/10.1103/PhysRevB.86.224411} {\bibfield  {journal} {\bibinfo
  {journal} {Phys. Rev. B}\ }\textbf {\bibinfo {volume} {86}},\ \bibinfo
  {pages} {224411} (\bibinfo {year} {2012})}\BibitemShut {NoStop}%
\bibitem [{\citenamefont {Haldane}(1985)}]{Haldane1985}%
  \BibitemOpen
  \bibfield  {author} {\bibinfo {author} {\bibfnamefont {F.~D.~M.}\
  \bibnamefont {Haldane}},\ }\bibfield  {title} {\bibinfo {title} {{$\Theta$
  physics and quantum spin chains}},\ }\href {https://doi.org/10.1063/1.335096}
  {\bibfield  {journal} {\bibinfo  {journal} {J. Appl. Phys.}\ }\textbf
  {\bibinfo {volume} {57}},\ \bibinfo {pages} {3359} (\bibinfo {year}
  {1985})}\BibitemShut {NoStop}%
\bibitem [{\citenamefont {Haldane}(1988)}]{Haldane1988}%
  \BibitemOpen
  \bibfield  {author} {\bibinfo {author} {\bibfnamefont {F.~D.~M.}\
  \bibnamefont {Haldane}},\ }\bibfield  {title} {\bibinfo {title} {{O(3)
  nonlinear $\sigma$ model and the topological distinction between integer- and
  half-integer-spin antiferromagnets in two dimensions}},\ }\href
  {https://doi.org/10.1103/PhysRevLett.61.1029} {\bibfield  {journal} {\bibinfo
   {journal} {Phys. Rev. Lett.}\ }\textbf {\bibinfo {volume} {61}},\ \bibinfo
  {pages} {1029} (\bibinfo {year} {1988})}\BibitemShut {NoStop}%
\bibitem [{\citenamefont {Tang}\ and\ \citenamefont
  {Sandvik}(2011)}]{Tang2011a}%
  \BibitemOpen
  \bibfield  {author} {\bibinfo {author} {\bibfnamefont {Y.}~\bibnamefont
  {Tang}}\ and\ \bibinfo {author} {\bibfnamefont {A.~W.}\ \bibnamefont
  {Sandvik}},\ }\bibfield  {title} {\bibinfo {title} {{Method to characterize
  spinons as emergent elementary particles}},\ }\href
  {https://doi.org/10.1103/PhysRevLett.107.157201} {\bibfield  {journal}
  {\bibinfo  {journal} {Phys. Rev. Lett.}\ }\textbf {\bibinfo {volume} {107}},\
  \bibinfo {pages} {157201} (\bibinfo {year} {2011})}\BibitemShut {NoStop}%
\bibitem [{\citenamefont {Tang}\ and\ \citenamefont
  {Sandvik}(2015)}]{Tang2015}%
  \BibitemOpen
  \bibfield  {author} {\bibinfo {author} {\bibfnamefont {Y.}~\bibnamefont
  {Tang}}\ and\ \bibinfo {author} {\bibfnamefont {A.~W.}\ \bibnamefont
  {Sandvik}},\ }\bibfield  {title} {\bibinfo {title} {{Quantum Monte Carlo
  studies of spinons in one-dimensional spin systems}},\ }\href
  {https://doi.org/10.1103/PhysRevB.92.184425} {\bibfield  {journal} {\bibinfo
  {journal} {Phys. Rev. B}\ }\textbf {\bibinfo {volume} {92}},\ \bibinfo
  {pages} {184425} (\bibinfo {year} {2015})}\BibitemShut {NoStop}%
\bibitem [{\citenamefont {Matsumoto}\ \emph {et~al.}(2001)\citenamefont
  {Matsumoto}, \citenamefont {Yasuda}, \citenamefont {Todo},\ and\
  \citenamefont {Takayama}}]{Matsumoto2001a}%
  \BibitemOpen
  \bibfield  {author} {\bibinfo {author} {\bibfnamefont {M.}~\bibnamefont
  {Matsumoto}}, \bibinfo {author} {\bibfnamefont {C.}~\bibnamefont {Yasuda}},
  \bibinfo {author} {\bibfnamefont {S.}~\bibnamefont {Todo}},\ and\ \bibinfo
  {author} {\bibfnamefont {H.}~\bibnamefont {Takayama}},\ }\bibfield  {title}
  {\bibinfo {title} {{Ground-state phase diagram of quantum Heisenberg
  antiferromagnets on the anisotropic dimerized square lattice}},\ }\href
  {https://doi.org/10.1103/PhysRevB.65.014407} {\bibfield  {journal} {\bibinfo
  {journal} {Phys. Rev. B}\ }\textbf {\bibinfo {volume} {65}},\ \bibinfo
  {pages} {014407} (\bibinfo {year} {2001})}\BibitemShut {NoStop}%
\bibitem [{\citenamefont {Ma}\ \emph {et~al.}(2018)\citenamefont {Ma},
  \citenamefont {Weinberg}, \citenamefont {Shao}, \citenamefont {Guo},
  \citenamefont {Yao},\ and\ \citenamefont {Sandvik}}]{Ma2018}%
  \BibitemOpen
  \bibfield  {author} {\bibinfo {author} {\bibfnamefont {N.}~\bibnamefont
  {Ma}}, \bibinfo {author} {\bibfnamefont {P.}~\bibnamefont {Weinberg}},
  \bibinfo {author} {\bibfnamefont {H.}~\bibnamefont {Shao}}, \bibinfo {author}
  {\bibfnamefont {W.}~\bibnamefont {Guo}}, \bibinfo {author} {\bibfnamefont
  {D.-X.}\ \bibnamefont {Yao}},\ and\ \bibinfo {author} {\bibfnamefont {A.~W.}\
  \bibnamefont {Sandvik}},\ }\bibfield  {title} {\bibinfo {title} {{Anomalous
  quantum-critical scaling corrections in two-dimensional antiferromagnets}},\
  }\href {https://doi.org/10.1103/PhysRevLett.121.117202} {\bibfield  {journal}
  {\bibinfo  {journal} {Phys. Rev. Lett.}\ }\textbf {\bibinfo {volume} {121}},\
  \bibinfo {pages} {117202} (\bibinfo {year} {2018})}\BibitemShut {NoStop}%
\bibitem [{\citenamefont {Sandvik}(2007)}]{Sandvik2007}%
  \BibitemOpen
  \bibfield  {author} {\bibinfo {author} {\bibfnamefont {A.~W.}\ \bibnamefont
  {Sandvik}},\ }\bibfield  {title} {\bibinfo {title} {{Evidence for deconfined
  quantum criticality in a two-dimensional Heisenberg model with four-spin
  interactions}},\ }\href {https://doi.org/10.1103/PhysRevLett.98.227202}
  {\bibfield  {journal} {\bibinfo  {journal} {Phys. Rev. Lett.}\ }\textbf
  {\bibinfo {volume} {98}},\ \bibinfo {pages} {227202} (\bibinfo {year}
  {2007})}\BibitemShut {NoStop}%
\bibitem [{\citenamefont {Lou}\ \emph {et~al.}(2009)\citenamefont {Lou},
  \citenamefont {Sandvik},\ and\ \citenamefont {Kawashima}}]{Lou2009a}%
  \BibitemOpen
  \bibfield  {author} {\bibinfo {author} {\bibfnamefont {J.}~\bibnamefont
  {Lou}}, \bibinfo {author} {\bibfnamefont {A.~W.}\ \bibnamefont {Sandvik}},\
  and\ \bibinfo {author} {\bibfnamefont {N.}~\bibnamefont {Kawashima}},\
  }\bibfield  {title} {\bibinfo {title} {{Antiferromagnetic to
  valence-bond-solid transitions in two-dimensional SU(N) Heisenberg models
  with multispin interactions}},\ }\href
  {https://doi.org/10.1103/PhysRevB.80.180414} {\bibfield  {journal} {\bibinfo
  {journal} {Phys. Rev. B}\ }\textbf {\bibinfo {volume} {80}},\ \bibinfo
  {pages} {180414(R)} (\bibinfo {year} {2009})}\BibitemShut {NoStop}%
\bibitem [{\citenamefont {Senthil}\ \emph
  {et~al.}(2004{\natexlab{a}})\citenamefont {Senthil}, \citenamefont {Balents},
  \citenamefont {Sachdev}, \citenamefont {Vishwanath},\ and\ \citenamefont
  {Fisher}}]{Senthil2004a}%
  \BibitemOpen
  \bibfield  {author} {\bibinfo {author} {\bibfnamefont {T.}~\bibnamefont
  {Senthil}}, \bibinfo {author} {\bibfnamefont {L.}~\bibnamefont {Balents}},
  \bibinfo {author} {\bibfnamefont {S.}~\bibnamefont {Sachdev}}, \bibinfo
  {author} {\bibfnamefont {A.}~\bibnamefont {Vishwanath}},\ and\ \bibinfo
  {author} {\bibfnamefont {M.~P.~A.}\ \bibnamefont {Fisher}},\ }\bibfield
  {title} {\bibinfo {title} {{Quantum criticality beyond the
  Landau-Ginzburg-Wilson paradigm}},\ }\href
  {https://doi.org/10.1103/PhysRevB.70.144407} {\bibfield  {journal} {\bibinfo
  {journal} {Phys. Rev. B}\ }\textbf {\bibinfo {volume} {70}},\ \bibinfo
  {pages} {144407} (\bibinfo {year} {2004}{\natexlab{a}})}\BibitemShut
  {NoStop}%
\bibitem [{\citenamefont {Senthil}\ \emph
  {et~al.}(2004{\natexlab{b}})\citenamefont {Senthil}, \citenamefont
  {Vishwanath}, \citenamefont {Balents}, \citenamefont {Sachdev},\ and\
  \citenamefont {Fisher}}]{Senthil2004b}%
  \BibitemOpen
  \bibfield  {author} {\bibinfo {author} {\bibfnamefont {T.}~\bibnamefont
  {Senthil}}, \bibinfo {author} {\bibfnamefont {A.}~\bibnamefont {Vishwanath}},
  \bibinfo {author} {\bibfnamefont {L.}~\bibnamefont {Balents}}, \bibinfo
  {author} {\bibfnamefont {S.}~\bibnamefont {Sachdev}},\ and\ \bibinfo {author}
  {\bibfnamefont {M.~P.~A.}\ \bibnamefont {Fisher}},\ }\bibfield  {title}
  {\bibinfo {title} {{Deconfined quantum critical points}},\ }\href
  {https://doi.org/10.1126/science.1091806} {\bibfield  {journal} {\bibinfo
  {journal} {Science}\ }\textbf {\bibinfo {volume} {303}},\ \bibinfo {pages}
  {1490} (\bibinfo {year} {2004}{\natexlab{b}})}\BibitemShut {NoStop}%
\bibitem [{\citenamefont {Sandvik}(2005)}]{Sandvik2005}%
  \BibitemOpen
  \bibfield  {author} {\bibinfo {author} {\bibfnamefont {A.~W.}\ \bibnamefont
  {Sandvik}},\ }\bibfield  {title} {\bibinfo {title} {{Ground state projection
  of quantum spin systems in the valence-bond basis}},\ }\href
  {https://doi.org/10.1103/PhysRevLett.95.207203} {\bibfield  {journal}
  {\bibinfo  {journal} {Phys. Rev. Lett.}\ }\textbf {\bibinfo {volume} {95}},\
  \bibinfo {pages} {207203} (\bibinfo {year} {2005})}\BibitemShut {NoStop}%
\bibitem [{\citenamefont {Sandvik}\ and\ \citenamefont
  {Evertz}(2010)}]{Sandvik2010a}%
  \BibitemOpen
  \bibfield  {author} {\bibinfo {author} {\bibfnamefont {A.~W.}\ \bibnamefont
  {Sandvik}}\ and\ \bibinfo {author} {\bibfnamefont {H.~G.}\ \bibnamefont
  {Evertz}},\ }\bibfield  {title} {\bibinfo {title} {{Loop updates for
  variational and projector quantum Monte Carlo simulations in the valence-bond
  basis}},\ }\href {https://doi.org/10.1103/PhysRevB.82.024407} {\bibfield
  {journal} {\bibinfo  {journal} {Phys. Rev. B}\ }\textbf {\bibinfo {volume}
  {82}},\ \bibinfo {pages} {024407} (\bibinfo {year} {2010})}\BibitemShut
  {NoStop}%
\bibitem [{\citenamefont {Deng}\ \emph {et~al.}(2005)\citenamefont {Deng},
  \citenamefont {Bl{\"{o}}te},\ and\ \citenamefont {Nightingale}}]{Deng2005}%
  \BibitemOpen
  \bibfield  {author} {\bibinfo {author} {\bibfnamefont {Y.}~\bibnamefont
  {Deng}}, \bibinfo {author} {\bibfnamefont {H.~W.~J.}\ \bibnamefont
  {Bl{\"{o}}te}},\ and\ \bibinfo {author} {\bibfnamefont {M.~P.}\ \bibnamefont
  {Nightingale}},\ }\bibfield  {title} {\bibinfo {title} {{Surface and bulk
  transitions in three-dimensional O(n) models}},\ }\href
  {https://doi.org/10.1103/PhysRevE.72.016128} {\bibfield  {journal} {\bibinfo
  {journal} {Phys. Rev. E}\ }\textbf {\bibinfo {volume} {72}},\ \bibinfo
  {pages} {016128} (\bibinfo {year} {2005})}\BibitemShut {NoStop}%
\bibitem [{\citenamefont {Mermin}\ and\ \citenamefont
  {Wagner}(1966)}]{Mermin1966}%
  \BibitemOpen
  \bibfield  {author} {\bibinfo {author} {\bibfnamefont {N.~D.}\ \bibnamefont
  {Mermin}}\ and\ \bibinfo {author} {\bibfnamefont {H.}~\bibnamefont
  {Wagner}},\ }\bibfield  {title} {\bibinfo {title} {{Absence of ferromagnetism
  or antiferromagnetism in one- or two-dimensional isotropic Heisenberg
  models}},\ }\href {https://doi.org/10.1103/PhysRevLett.17.1133} {\bibfield
  {journal} {\bibinfo  {journal} {Phys. Rev. Lett.}\ }\textbf {\bibinfo
  {volume} {17}},\ \bibinfo {pages} {1133} (\bibinfo {year}
  {1966})}\BibitemShut {NoStop}%
\bibitem [{\citenamefont {Witten}(1984)}]{Witten1984}%
  \BibitemOpen
  \bibfield  {author} {\bibinfo {author} {\bibfnamefont {E.}~\bibnamefont
  {Witten}},\ }\bibfield  {title} {\bibinfo {title} {{Non-abelian bosonization
  in two dimensions}},\ }\href {https://doi.org/10.1007/BF01215276} {\bibfield
  {journal} {\bibinfo  {journal} {Commun. Math. Phys.}\ }\textbf {\bibinfo
  {volume} {92}},\ \bibinfo {pages} {455} (\bibinfo {year} {1984})}\BibitemShut
  {NoStop}%
\bibitem [{\citenamefont {Affleck}(1985{\natexlab{a}})}]{Affleck1985b}%
  \BibitemOpen
  \bibfield  {author} {\bibinfo {author} {\bibfnamefont {I.}~\bibnamefont
  {Affleck}},\ }\bibfield  {title} {\bibinfo {title} {{Critical behavior of
  two-dimensional systems with continuous symmetries}},\ }\href
  {https://doi.org/10.1103/PhysRevLett.55.1355} {\bibfield  {journal} {\bibinfo
   {journal} {Phys. Rev. Lett.}\ }\textbf {\bibinfo {volume} {55}},\ \bibinfo
  {pages} {1355} (\bibinfo {year} {1985}{\natexlab{a}})}\BibitemShut {NoStop}%
\bibitem [{\citenamefont {Affleck}(1985{\natexlab{b}})}]{Affleck1985a}%
  \BibitemOpen
  \bibfield  {author} {\bibinfo {author} {\bibfnamefont {I.}~\bibnamefont
  {Affleck}},\ }\bibfield  {title} {\bibinfo {title} {{The quantum Hall
  effects, $\sigma$-models at $\Theta$ = $\pi$ and quantum spin chains}},\
  }\href {https://doi.org/10.1016/0550-3213(85)90353-0} {\bibfield  {journal}
  {\bibinfo  {journal} {Nucl. Phys. B}\ }\textbf {\bibinfo {volume} {257}},\
  \bibinfo {pages} {397} (\bibinfo {year} {1985}{\natexlab{b}})}\BibitemShut
  {NoStop}%
\end{thebibliography}%

\appendix

\setcounter{table}{0}
\renewcommand{\thetable}{A\Roman{table}}
\setcounter{equation}{0}
\renewcommand{\theequation}{A\arabic{equation}}

\section{Extraordinary-log  trial fitting in the $Q<Q_{c}$ phase}

\begin{ruledtabular}
\begin{table}[tb]
\caption{Fitting parameters of $(-1)^{L/2}C_{\parallel}(L/2)$ at $Q=0$ according to Eq.~(\ref{eq:exlog1}). The fitting quality is very poor. In addition, the parameters vary significantly with the minimum lattice size $L_{\mr{min}}$, indicating that the fitting is unstable.}
\label{tab:exlog1}
\begin{tabular}{c c c c c}
$L_{\mr{min}}$	& $a$						& $q_{\parallel}$	& $c$			& $\chi$/d.o.f.	\\
\hline
12				& $1.2(9)\times10^{7}$		& 7.6(2)			& 8.9(3)		& 35.2			\\
18				& $(2.7\pm1.2)\times10^{3}$	& 5.09(15)			& 4.76(24)		& 13.9			\\
24				& $(3.4\pm4.7)\times10^{5}$	& 6.6(4)			& 7.2(7)		& 11.0			\\
30				& $(5.6\pm5.2)\times10^{2}$	& 4.6(3)			& 3.9(5)		& 6.7			\\
36				& $(0.8\pm3.8)\times10^{7}$	& $7.5\pm1.3$		& $8.9\pm2.3$	& 3.9			\\
42				& $(3.3\pm7.1)\times10^{2}$	& 4.4(7)			& $3.6\pm1.2$	& 2.1			\\
48				& $(0.3\pm2.4)\times10^{6}$	& $6.5\pm2.5$		& $7.2\pm4.4$	& 2.0
\end{tabular}
\end{table}
\end{ruledtabular}

\begin{ruledtabular}
\begin{table}[tb]
\caption{Fitting parameter $q_{\parallel}$ of $S(L)$ at $Q=0$ according to Eq.~(\ref{eq:exlog2}) obtained with different maximal numbers of iteration. Its monotonic variation indicates that the fitting does not converge at all. The precision goal in the fitting procedure is set to be $10^{-10}$.}
\label{tab:exlog2}
\begin{tabular}{c c c c}
$L_{\mr{min}}$	& $q_{\parallel}$ ($10^{3}$ iter.)	& $q_{\parallel}$ ($10^{4}$ iter.)	& $q_{\parallel}$ ($10^{5}$ iter.)	\\
\hline
12				& 8.55(9)							& 20.7711(10)						& 30.8671(14)						\\
18				& 7.54(12)							& 18.1538(11)						& 26.1377(15)						\\
24				& 6.64(15)							& 15.6926(13)						& 20.1417(16)						\\
30				& 6.10(20)							& 14.0463(16)						& 17.4503(18)						\\
36				& 5.58(26)							& 12.9260(19)						& 17.7581(24)						\\
42				& 5.0(3)							& 11.2666(23)						& 13.6221(26)						\\
48				& 4.6(4)							& 10.3580(28)						& 14.316(4)
\end{tabular}
\end{table}
\end{ruledtabular}

We attempt to fit the boundary spin correlation function in the $Q<Q_{c}$ phase with the extraordinary-log scaling form~\cite{Metlitski2022, ParisenToldin2021},
\begin{equation}
(-1)^{L/2}C_{\parallel}(L/2) = a(\ln r+c)^{-q_{\parallel}},
\label{eq:exlog1}
\end{equation}
and the corresponding static AF structure factor with
\begin{equation}
S(L)=L^{-1}\sum_{r=0}^{L-1}(-1)^{r}C_{\parallel}(r) = a'(\ln L+c')^{-q_{\parallel}},
\label{eq:exlog2}
\end{equation}
where $a$, $a'$, $c$, $c'$, and $q_{\parallel}$ are fitting parameters. The trial fitting results at $Q=0$ are listed in Tables~\ref{tab:exlog1} and \ref{tab:exlog2}. The fitting quality of $(-1)^{L/2}C_{\parallel}(L/2)$ is very poor. In addition, the fitting parameters change significantly when data of the smallest lattice sizes are progressively excluded in the fitting, indicating that the fitting is unstable. Moreover, the fitting of $S(L)$ does not even converge: The parameters vary monotonically with the maximum number of iteration, indicating that the fitting algorithm fails to locate the stable minimum. Therefore, the extraordinary-log scaling form cannot adequately describe our numerical results in the $Q<Q_{c}$ phase.

\end{document}